\documentstyle[twoside,fleqn,espcrc2,epsf]{article}

\newcommand{\AmS}{{\protect\the\textfont2
  A\kern-.1667em\lower.5ex\hbox{M}\kern-.125emS}}

\newcommand{\al}{\alpha}
\newcommand{\bt}{\beta}
\newcommand{\gm}{\gamma}

\newcommand{\lm}{\lambda}

\newcommand{\ph}{\phi}
\newcommand{\vr}{\varphi}

\newcommand{\Gm}{\Gamma}

\newcommand{\eela}[1]{\label{#1}\end{equation}}
\newcommand{\eeala}[1]{\label{#1}\end{eqnarray}}
\newcommand{\be}{\begin{eqnarray}}
\newcommand{\ee}{\end{eqnarray}}
\newcommand{\bea}{\begin{eqnarray}}
\newcommand{\eea}{\end{eqnarray}}
\title{Sphaleron rate at high temperature in 1+1 dimensions}

\author{Jan Smit
        and 
        Wai Hung Tang\address{Institute of Theoretical Physics, 
        University of Amsterdam \\
        Valckenierstraat 65, 1018 XE Amsterdam, the Netherlands}
\thanks{Supported by FOM}}
       
\begin{document}

\begin{abstract}
We resolve the controversy in the high temperature behavior of the
sphaleron rate in the abelian Higgs model in 1+1 dimensions.
The $T^2$ behavior at intermediate lattice spacings is found
to change into $T^{2/3}$ behavior in the continuum limit.
The results are supported by analytic arguments that the classical 
approximation is good for this model.
\end{abstract}

\maketitle

Sphaleron physics plays an important role in theories of
baryogenesis.  A simple but useful model is the 
abelian Higgs model in 1+1 dimensions, given by
\bea 
S&=&-\int {\rm d}^2 x\; \left[ \frac{1}{4g^2} F_{\mu\nu}F^{\mu\nu} +
(D_\mu \phi)^\ast D^\mu \phi
\right. \nonumber\\ &&\left. \mbox{} 
 + \mu^2 |\phi|^2 +\lambda |\phi|^4
\right], 
\nonumber
\eea
Recall that in 1+1 dimensions 
$\sqrt{\lambda}$ and $g$ have dimension of mass.
A toy model for the electroweak theory is obtained by coupling to
fermions, such that
in the quantum theory the fermion current is anomalous. 
Changes in fermion number are then 
proportional to changes in Chern-Simons number 
\be
C = \frac{1}{2\pi}\int_0^L {\rm d}x_1\; A_1,
\nonumber
\ee
thereby mimicking the B+L violation in the Standard Model.
Here we assume space to be a circle of circumference $L$.

The sphaleron rate (of fermion number violation)
$\Gamma$ can be identified from the diffusion of
Chern-Simons number 
\be
\Gamma =\frac{1}{t} \left\langle \left[ C(t)-C(0) \right]^2
\right\rangle, \hspace{5mm} t\rightarrow \infty.
\nonumber
\ee
For relatively low temperatures this rate is exponentially suppressed
by the sphaleron barrier. 
Numerical simulations 
\cite{FoKraPo94,SmTa94,SmTa95}
aggree with analytic results \cite{BochTsi89}
in this regime. 
At high temperature the rate is not known analytically, but expected
to be un-suppressed. For temperatures larger than any mass
scale one may naively expect on dimensional grounds
\be
\frac{\Gm}{L} \propto T^2 \label{b1}.
\ee
Such behavior was indeed found by us numerically
and reported at LATTICE 94 \cite{SmTa95}.
However, De Forcrand, Krasnitz and Potting \cite{FoKraPo94} gave a scaling 
argument that the behavior should instead be given by
\be
\frac{\Gm}{L} \propto T^{2/3}. \label{b2}
\ee
\begin{figure}[t]
\epsfxsize 7.5cm
\centerline{\epsfbox{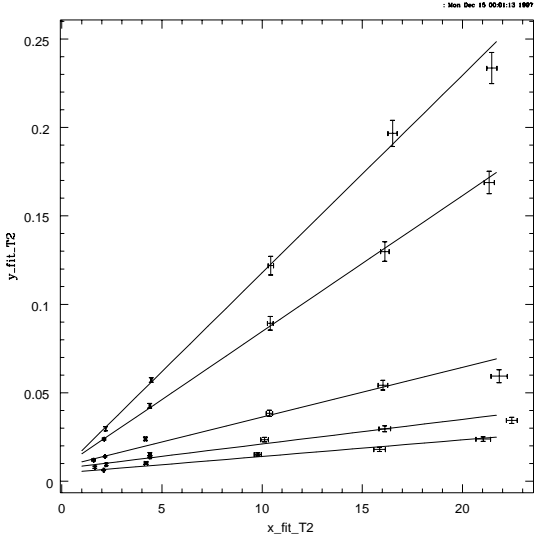}}
\caption{New data for $F= \Gm/(m_{\ph}^2 L)$
versus $T'^2$ for the lattice spacings
given in \protect (\ref{ahs}).
The straight lines are fits to the data.
The lower three lines do not fit the data very well.
}
\label{fig1}
\end{figure}

This behavior (\ref{b2}) 
can also be argued for as follows.
The simulations use the classical approximation 
\bea
&&\langle [C(t)-C(0)]^2 \rangle \approx \mbox{} \nonumber\\
&&
\frac{ \int_{\vr,\pi} e^{-H_{\rm eff}(\vr,\pi)/T}\,
\left[ C(\vr(t),\pi(t)) - C(\vr,\pi)\right]^2 }
{\int_{\vr,\pi} e^{-H_{\rm eff}(\vr,\pi)/T} }.
\nonumber
\label{defclas}
\eea
Here $\vr$ and $\pi$ denote generic canonical variables and
$\vr(t)$ and $\pi(t)$ are solutions of the classical Hamilton
equations with initial conditions
$\vr(0)=\vr$, $\pi(0)=\pi$. The effective hamiltonian $H_{\rm eff}$
is approximated by its classical form. Rescaling $\vr\to\vr\sqrt{T}$,
$\pi\to\pi\sqrt{T}$ produces $T$ only in the combination $\lm T$
and $e^2 T$.
Since this combination has mass
dimension three, the behavior (\ref{b2}) in the form 
$\Gm/L \propto (\lm T)^{2/3}$ appears natural.

We analyzed the quality of the classical approximation in perturbation theory
and found the following favorable properties (in 1+1 dimensions!) 
\cite{TaSm98}:\\
\begin{itemize}
\item 
Correlation functions of the basic fields are finite. 
\item
The approximation becomes exact in the weak coupling/high temperature limit
\bea
\lm &=& v^{-2} |\bar\mu|^2,\;\;\;
g^2 = \xi \lm,\;\;\;
T = v^2 |\bar\mu| T',\;\;\;
\nonumber\\
v^2 &\to& \infty, 
\;\;\; \xi,T'\;\; \mbox{fixed}.
\label{limit}
\eea
Here $\bar\mu$ is the renormalized mass parameter in the 
$\overline{\mbox{MS}}$ scheme and 
$v^2 = -\mu^2/\lm$
is the classical ground state value of $2\vr^*\vr$.
In the limit (\ref{limit}), $\mu^2/\bar\mu^2 \to 1$.
Notice that  
$T' = \lm T/|\bar\mu|^3$
involves the combination $\lm T$.
\end{itemize}
In our previous results \cite{SmTa95} the 
coefficient of $T^2$ appeared to vanish on extrapolation
of the lattice distance to zero. At the time we interpreted this
as an effect caused by using a too simple effective hamiltonian
(the classical one), but
now the perturbative analysis tells us that using the classical hamiltonian
is fine in the limit (\ref{limit}).
To settle the issue we carried out additional simulations at higher 
temperatures and smaller lattice spacings. 

Fig.\ 1 shows the dimensionless rate 
\be
F \equiv \frac{\Gm}{m_{\ph}^2 L}
\nonumber
\ee
plotted versus $T'^2$, for several lattice spacings, 
\be
a|\mu| = 0.25,\; 0.23,\; 0.16,\; 0.11,\; 0.08. \label{ahs}
\ee
There is clear $T^2$ behavior for the two larger spacings,
but for the three 
smaller spacings this  behavior does not fit the data any more.
A crossover appears to take place between $a|\mu| = 0.23$ and 0.16.
In fact,
$T^{2/3}$ behavior fits the data better at the three smaller spacings. 
Figs.\ 2--3 illustrate
the behavior in more quantitative detail.
Fitting the forms 
\bea
F_{2} &=& c_0 + c_2 T'^2, \nonumber\\
F_{2/3} &=& c_0 + c_{2/3} T'^{2/3}, \nonumber
\eea
led to 
\bea
\chi_{2}^2/\mbox{d.o.f.} &=& 1.4,\; 0.42,\; 6.5,\; 4.8,\; 3.45,\nonumber\\
\chi_{2/3}^2/\mbox{d.o.f.} &=& 28.9,\; 21.6,\; 2.5,\; 1.1,\; 1.2,\nonumber
\eea
for
$a|\mu| = 0.25$, 0.23, 0.16, 0.11, 0.08, respectively.
The volume was fixed at $|\mu| L = 16$, and $\xi = 4$.
 
\begin{figure}[hb]
\epsfxsize 7.5cm
\centerline{\epsfbox{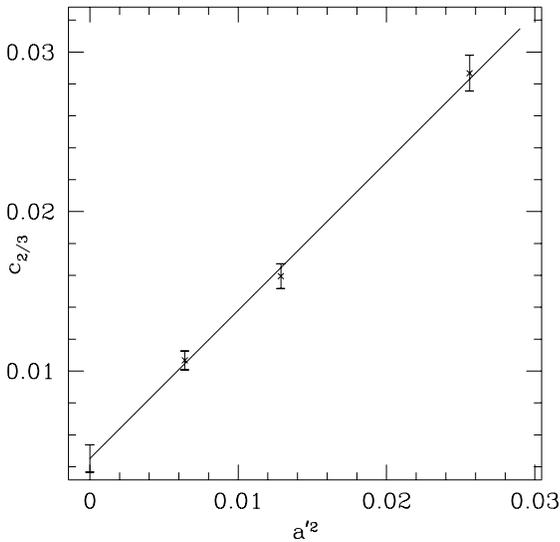}}
\caption{Plot of $c_{2/3}$ as a function of $a^2 |\mu|^2$.
}
\label{fato0}
\end{figure}

Clearly,
the $T^{2/3}$ form is favored for the three smaller lattice spacings.
Extrapolating the resulting $c_{2/3}$ to zero lattice spacing, 
assuming a quadratic dependence on $a$ (cf.\ Fig.\ \ref{fato0}), gives
the result $c_{2/3} = 0.00452 (86)$ for $a=0$,
which translates into
\be 
\frac{\Gm}{L} = 0.0090(17) (\lm T)^{2/3}, \;\;\; g^2/\lm = 4.
\nonumber
\ee
Arguments for a
 $T^{2/3}$ behavior were given in ref.\ \cite{FoKraPo94}, where it 
was shown that the classical rate could be written in
terms of a function $g(x,y)$, as
$\Gm/L = T^{2/3} g(a^3 T, v^3/T)$,
using units in which $\lm = 1$, suppressing $\xi$ dependence.
For high temperatures  $T^{2/3}$ behavior then followed
in the continuum limit provided $g(0,0) \neq 0, \infty$ existed.
For the case $v^2=0$, numerical data supported a nontrivial $g(0,0)$.
Our analytic results \cite{TaSm98} support the existence of a
continuum limit of $g(a^3 T,v^3/T)$, but notice how the
combination $a^3 T$ implies
non-commutativity of the limits $a\to 0$ and $T\to\infty$.
Unfortunately, our attempts to extrapolate first to zero lattice spacing 
failed because the resulting errors got too large. Our present
analysis seems to indicate a behavior like 
$g(x,y) = g(0,0) + \bt (xy)^{2/3} + \al x^{4/3} + \gm y^{4/3} + \cdots$,
or $\Gm/L = [g(0,0) + \bt a^2] T^{2/3} + \al a^4 T^2 + \gm T^{-2/3} + \cdots$.

\begin{figure}[t]
\epsfxsize 7.5cm
\epsfbox{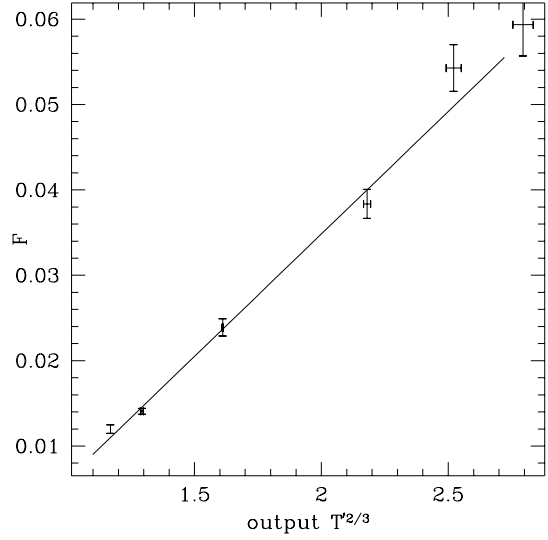}
\epsfxsize 7.5cm
\epsfbox{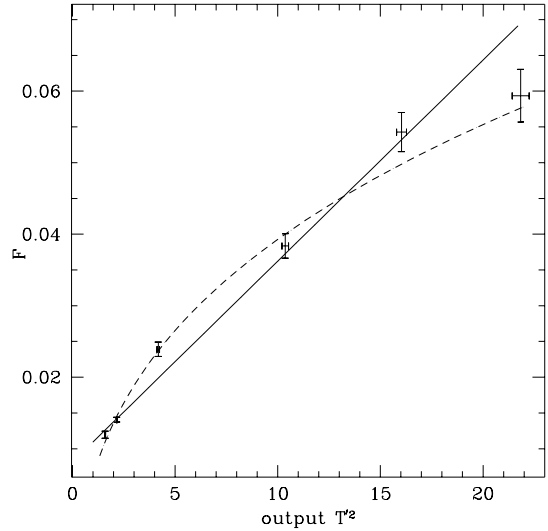}
\caption{Top: data showing $T^{2/3}$ behavior for $a|\mu|=0.16$.
Plotted is $F$ versus $T'^{2/3}$ and a fit to $c_0 + c_{2/3} T'^{2/3}$.
Bottom:
same data plotted versus $T'^2$.
The straight line is a fit of $c_0 + c_{2} T'^{2}$.
The curved line is the $T'^{2/3}$ fit of the top figure.
The data lack $T^2$ behavior but favor $T'^{2/3}$ behavior
in the region $a|\mu| \leq 0.16$.
}
\label{fig3a}
\end{figure}

\end{document}